\documentclass[12pt]{article}
\usepackage{graphicx}
\newcommand{\sss}{\scriptscriptstyle}

\newcommand {\be}{\begin{equation}} 
\newcommand{\ee}{\end{equation}}    

\def\dds1{\frac{\partial}{\partial s_1}}

\def\vti{v_{{\sss T}i}}
\def\vte{v_{{\sss T}e}}

\def\d{d\kern-0.8 ex\vrule height 1.3 ex depth-1.24 ex width 0.7 ex
\kern 0.15 ex}
\def\D{D\kern-1.7 ex\vrule height .87 ex depth-0.8 ex width 0.7 ex
\kern 0.95 ex}
\def\nabp{\nabla_{\perp}}

\begin{document}
\baselineskip 20 pt

\begin{center}

\Large{\bf Solar nanoflares and other smaller energy release events as growing drift waves}

\end{center}

\vspace{0.7cm}

\begin{center}

 J. Vranjes and S. Poedts

Center for Plasma Astrophysics, and Leuven Mathematical Modeling and Computational Science Centre
 (LMCC),  Celestijnenlaan 200B, 3001 Leuven,  Belgium.

\end{center}

\vspace{2cm}

{\bf Abstract:}

Rapid energy releases (RERs) in the solar corona extend over many
orders of magnitude, the largest (flares) releasing an energy of
$10^{25}\;$J or more. Other events, with a typical energy that is a
billion times less, are called nanoflares.  A basic difference
between flares and nanoflares  is that flares need a larger magnetic
field and thus occur only in active regions, while nanoflares can
appear everywhere. The origin of such RERs  is usually attributed to
magnetic reconnection that takes place at altitudes just above the
transition region. Here we show that nanoflares and smaller similar
RERs can be explained within the drift wave theory as a natural
stage in the kinetic  growth of the drift wave. In this scenario, a
growing mode with a sufficiently large amplitude leads to stochastic
heating that can provide an energy release of over  $10^{16}\;$J.


\pagebreak


\noindent{\bf I. \,\,\, INTRODUCTION}

\vspace{0.5cm}

The central problems regarding rapid energy releases (RERs) are explaining i)~where the
energy comes from, ii)~by what mechanism  it is released, and
iii)~how it is transformed into (random and directed) particle
energy. A widely (but not universally) used concept in the
explanation of flares and other RERs, is that of magnetic
reconnection which implies the presence of electric currents. Yet,
the origin, and consequently the source of these currents, remains
not well understood. Most frequently, shearing photospheric motions
and/or magnetic flux emergence  are  used \cite{wh} in the models.
The concept has been questioned in the past \cite{jan,pud} because
of lack of a measurable magnetic energy and configuration change in
some  events \cite{jan,may}.

 The electric field associated with RER
events has been studied extensively in the past \cite{dav,zh,be,fo},
with very strong field reported \cite{dav} of $7\cdot 10^4\;$V/m ,
that may go \cite{zh} up to $1.3 \cdot 10^5\;$V/m. As a rule, its
presence is associated with the acceleration of lighter particles.
Apart from the particle acceleration, flares and other smaller
energy release events are responsible for strong local heating
\cite{as}. However, within the current models, it is nontrivial to
explain how the released energy is dissipated so fast in the highly
conductive corona, characterized by a Lundquist number around
$10^{13}$. The existing continuum approximation (fluid) or
magnetohydrodynamics (MHD) models, are not very promising because it
is clear that the actual heating takes place at length scales much
smaller than those on which the (macroscopic) MHD model is
justified. We stress also that the possibility can not be excluded
that, at least in some cases, the reconnection can appear as a
consequence, rather than a cause of a RER. Another problematic
aspect regarding the role of RERs in the heating, is the
distribution of magnetic null points. According to Ref. \cite{reg}
only 2\% of them are located in the corona and 54\% in the
photosphere. This  is opposite to what would be required for the
mechanism that is presently believed  to heat the corona. The
heating by waves rather than by reconnection is also supported by
the diagnostic of active regions presented in Ref. \cite{mil}.

Here we present an alternative model of RERs and the consequent
plasma heating, based on the kinetic theory of drift waves, driven
by density gradients that are omnipresent in the solar corona. The
model is able to provide answers to all the fundamental questions
raised above. The required density gradients are visible in the
magnetic loops spread throughout the solar atmosphere. They are
closely connected to the continuous restructuring of the magnetic
field (due to the frozen-in conditions), and also to  the observed
inflow of the plasma along the magnetic loops \cite{sh}.
Measurements by Voyager~1 and Voyager~2 show \cite{woo,w} that the
transverse size of some of these highly elongated structures at the
Sun can be below 1$\;$km. Hinode observations \cite{8} only confirm
that the solar atmosphere is a highly structured and inhomogeneous
system. Moreover, a three-dimensional analysis \cite{as2} of the
coronal loops reveals short-scale density irregularities within each
loop separately.  Presently, the observable characteristic
dimensions of density irregularities are limited  by the available
resolution of the instruments (at best about $0.5\;$arcsecond).
However, even extremely short, meter-size scales can not be
excluded. This can be seen by calculating the perpendicular (to the
magnetic field) diffusion coefficient  for a corona environment
$D_{\bot, j}\approx \kappa T_j \nu_j/(m_j \Omega_j^2)\propto
m_j^{1/2}$, $j=i, e$. The diffusion velocity in the direction of the
given density gradient is \cite{v3a} $D_{\bot,j} \nabla n_0/n_0$.
Taking the inhomogeneity scale-length $L_n\equiv
[(dn_0/dx)/n_0]^{-1}=10, \,10^2,\, 10^5\;$m, where $x$-denotes the
direction perpendicular to the magnetic field vector,  we obtain the
ion diffusion velocities, respectively, $10^{-3}, 10^{-4},
10^{-7}\;$m/s only. Therefore, even very short density
inhomogeneities can last long enough to support relatively high
frequency drift instabilities. In dealing with the drift wave, we
may thus operate with  the density inhomogeneity scale lengths that
can have any value from meter size up to thousands of kilometers in
the case of coronal plumes. The role of the drift wave in RERs has
been overlooked so far in the literature, probably due to the fact
that it simply does not exist in the widely used MHD model.

The purpose of this work is to show that the free energy stored in
these density gradients may drive drift waves, and these may further
release energy  on a massive scale. The dissipation of the drift
waves is easy to explain in our self-consistent kinetic model that
works on the (very small) length scales at which the actual
dissipation takes place. Actually, two mechanisms of energy exchange
and heating will be shown to take place simultaneously, one due to
the Landau effect and another one, {\em stochastic heating} which acts exclusively in the perpendicular
direction.  To prove this we use only established basic theory,
verified experimentally in laboratory plasmas.

\vspace{0.5cm}

\noindent{\bf II. \,\,\, MODEL AND RESULTS}

\vspace{0.5cm}

 The drift wave
properties within well known limits  are described \cite{is} by the
frequency and the growth rate, respectively
 \be
\omega_r=-\frac{\omega_{*i} \Lambda_0(b_i)}{1- \Lambda_0(b_i) +
T_i/T_e + k_y^2 \lambda_{di}^2},\label{k1} \ee
\[
\gamma \simeq \left(\frac{\pi}{2}\right)^{1/2} \frac{\omega_r^2}
{\omega_{*i} \Lambda_0(b_i)}\left[\frac{T_i}{T_e} \frac{\omega_r -
\omega_{*e}}{|k_z|\vte} \exp[-\omega_r^2/(k_z^2 \vte^2)]\right.
\]
\be \left. +  \frac{\omega_r - \omega_{*i}}{|k_z|\vti}
\exp[-\omega_r^2/(k_z^2 \vti^2)]\right]. \label{k2} \ee
Here, $ \Lambda_0(b_i)=I_0(b_i) \exp(-b_i)$,  $b_i=k_y^2 \rho_i^2$,
$\rho_i = \vti/\Omega_i$, $  \lambda_{di}=\vti/\omega_{pi}$,
$\omega_{*i}=- \omega_{*e} T_i/T_e$, $\omega_{*e}=k_y v_{*e}$, $\vec
v_{*e}=-(v_{{\sss T} e}^2/\Omega_e) \vec e_z \times \nabp n_0/n_0$,
 $I_0$ is the modified Bessel function of the first kind and of
the order 0,  $\vec B_0=B_0 \vec e_z$, and $\nabla_\bot=\vec e_x
d/dx$.

In Eq.~(\ref{k2}) the positive  second  term in the square bracket
describes the damping on ions. The necessary condition for the
instability  $\omega_r<\omega_{*e}$, from the first term is, as a
rule, easily satisfied. Note that  Eq.~(\ref{k1}) reveals the
presence of the energy source already in the real part of the
frequency $\omega_r\propto \nabla_\bot n_0$, while details of its
growth due to the same source are described by Eq.~(\ref{k2}).

In Fig.~1  we give the wave growth-rate as a function of two parameters,
the perpendicular wavelength and  the density scale-length $L_n $.
 The chosen parameters correspond to the inner corona, i.e.,
 $B_0=5\cdot 10^{-3}\;$T, $n_0=10^{16}\;$m$^{-3}$, $T_i=T_e=T_0=7\cdot
10^{5}\;$K.  The graph does not
change drastically  by varying these parameters for $\pm 1$ order of
magnitude.  The wavelength parallel to the magnetic field vector  is taken as $\lambda_z= s\cdot 10^4\;$m.
We have introduced a parameter $s$ that we can vary within the range $10^{-1} - 10^3$ in order to demonstrate that the mode behavior remains unchanged at various density
scale lengths $L_n$, provided  that we keep the ratio $\lambda_z/L_n$ constant. It can easily be shown that
in this case the ratio $\gamma/\omega_r$ also remains constant \cite{vmnras,vepl}, while both quantities are shifted either to lower or higher values.
As example, the mode frequency for $s=1$,
$L_n=100\;$m, and $\lambda_y=1\;$m is $\omega_r=231\;$Hz, and it
changes as $231/s$ when the other parameters are fixed.
 The growth rate may easily become larger than $\omega_r$ for short
$\lambda_y, L_n$.

\begin{figure}
\includegraphics[height=8cm, bb=50 15 293 234, clip=,width=.6\columnwidth]{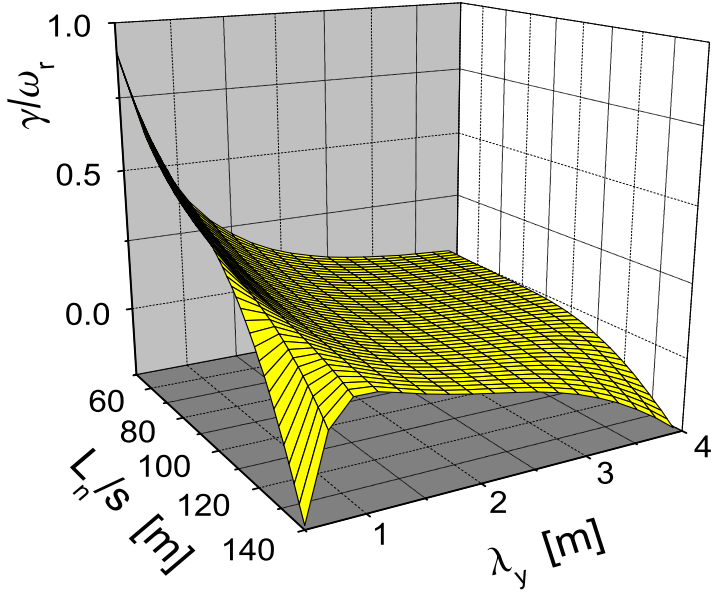}
\caption{\label{fig1} The growth rate (\ref{k2}) normalized to the
wave frequency $\omega_{r}$  in terms of the perpendicular
wavelength and  the density scale-length $L_n $. }
\end{figure}

The growth of the wave results in  a stochastic heating mechanism that
implies  single particle interaction with  the wave. The  process
has been   experimentally verified \cite{san2,san}. In the
drift wave, the ions move in the perpendicular direction to large
distances and they feel the time-varying field of the wave due to
the polarization drift $ \vec v_p=(d\vec E/dt)/(\Omega_i B_0)$, and as a result their motion becomes stochastic.
This stochastic heating is anisotropic, acting  mainly in the
direction normal to the magnetic field $B_0$. The perpendicular
heating in the experiment was larger by about a factor 3 compared to
the parallel one \cite{san2,san}, and it  is exceptionally fast.  It
the same time, in view of the mass difference and  the given
physical picture, this heating scenario predominantly acts on the
ions, with  heavier ions more efficiently heated than lighter ones.
For drift wave electrostatic potential  perturbation of the form $\phi(x) \sin(k_y y + k_z z
- \omega t)$, $|k_y|\gg |k_z|$, one finds the ion polarization drift
velocity following the procedure from Ref.  \cite{bel}. For large enough particle displacements (large wave amplitudes) $\vec v_p=
\vec e_y [k_y \phi/(\Omega_i B_0)] (d/dt)\sin( k_y y - \omega t)=
\vec e_y
[ k_y \phi/(\Omega_i B_0)] (k_y dy/dt - \omega) \cos (k_y y - \omega t)$, where $dy/dt=v_p$. This yields
\be
\vec v_p=-\vec e_y \frac{\omega  k_y \phi}{\Omega_i B_0}
\frac{\cos(k_y y-\omega t)}{1- a \cos(k_y y-\omega t)}, \label{cc1}
\ee
\[
a= k_y^2 \rho_i^2 e \phi/(\kappa T_i).
\]
 It has been shown \cite{san}
that the  stochastic heating takes place for a large enough  wave
amplitude, more precisely for $a \geq 1$.
This condition implies  that  the ion displacement due to the polarization drift
is comparable to the perpendicular wavelength. This is because $\vec  v_{p}= ( \vec e_z\times \partial \vec  v_{\bot}/\partial t)/\Omega_i$, and $\vec  v_{\bot}$ is the leading order
$\vec E\times \vec B$-drift, so that $v_{p}\sim a \omega/k_y$ and the perpendicular displacement due to the polarization drift
is $\delta=v_{p}/\omega=a/k_y$. Another important feature
is that $\vec  v_{p} \sim \vec k_y$, hence the stochastic heating is due to the electrostatic property of the wave.

According to \cite{san2,san}, the maximum achieved bulk
ion velocity, proportional to the wave amplitude, is given by
\be
v_m\simeq [k_y^2 \rho_i^2 e \phi/(\kappa T_i) + 1.9]\Omega_i/k_y.
\label{vm} \ee
As  explained in Ref. \cite{san}, the factor $1.9$ appears after making Pincar\'{e} plots of particle trajectories for different values of $a$.
Slightly different values of this factor, i.e., 1.5 and 2.3, are reported in
Refs. \cite{dra,san3},   respectively.

Eq. (\ref{vm})  implies an effective increase in the temperature $T_{ef}=m_i
v_m^2/(3 \kappa)$, while the volumetric  energy increase of the
stochastically heated particles is $\Sigma_m=n_0 m_i
v_m^2\;$ [J/m$^3$], and the energy release rate is
$\Gamma_m=\Sigma_m/\tau_g\;$ [J/(m$^3$s)], where $\tau_g=1/\gamma$.

Equation~(\ref{vm}) reveals a minimum of $v_m$ at $\lambda_y=2 \pi
\rho_i [e \phi/(\kappa T_i)]^{1/2}$. This is seen also in Fig.~2. Here,  we fix $\phi=10^3\;$V and use  the same parameters as in
Fig.~1.

\begin{figure}
\includegraphics[height=8cm, bb=15 15 286 222, clip=,width=.6\columnwidth]{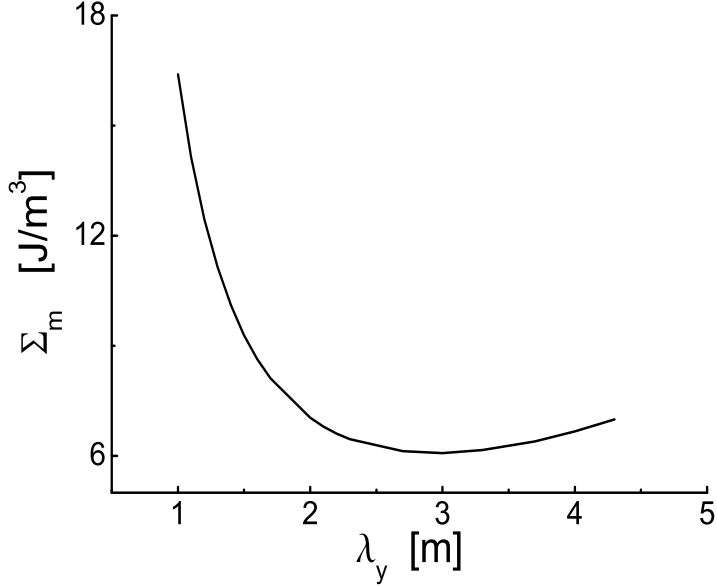}
\caption{\label{fig2} Volumetric energy released during stochastic
heating by the drift wave in terms of the perpendicular wavelength.
 }
\end{figure}

 Since  the results are valid
for any $s$, the growth time $\tau_g\propto s$ may have rather
different values. Assuming a small accidental initial perturbation $e
\phi_0/(\kappa T_i)=0.01$ the growth time till it gets the value
used above is $\tau_g=\log (\phi/\phi_0)/\gamma$.   Taking $s=1$ and
$s=10^3$, at the minimum from Fig.~2 we obtain $\tau_g \simeq
0.2\;$s, and $\tau_g \simeq 3.3\;$min, respectively.   The same
dependence on $s$ holds for the total plasma volume $V_p=L_x L_y
L_z$  involved in RERs, and consequently for the total amount of the
released energy. We may take  $L_x=L_n$, and for $L_z$ take one
wavelength $ \lambda_z$ only while for $L_y$ we may take a layer of
around $10\;$km. At the minimum $\lambda_y$ from Fig.~2 and for
$s=10^3$, this yields $\Sigma_m V_p=1.5 \cdot 10^{16}\;$J. Clearly,
this value can easily become considerably larger (e.g., for a larger
volume, then for $\lambda_y$ not taken at minimum,  and also
increasing the  density for one order of magnitude will increase the
energy for one order too).

Calculating $dT_{ef}(m_i)/d m_i$ it can
be shown that the heating increases with ion mass if $k_y^4 \rho_i^4
[e \phi/(\kappa T_i)]> 1.9$. For the parameters of interest here, it
turns out that this condition is always satisfied. We stress that
 numerous indications and observations in the solar corona \cite{hans,cus,cr3}
confirm these peculiar features of a stronger heating of heavier particles, i.e.,  $T_e<T_H<T_{He}$,
 and a stronger heating in the perpendicular direction as compared to the heating along the magnetic field vector.

The  drift mode presented above  implies a time-varying electric
field. Its parallel component $|\nabla_z \phi|$ for
$\lambda_z=s\cdot 10\;$km (in units of the Dreicer runaway electric
field) is $4$  and $0.004$ for $s=1$ and $10^3$, respectively. In
the first case, the bulk plasma species (primarily electrons) can be
accelerated by the wave in the $z$-direction. In the latter,  this
holds only for electrons from the tail in the distribution function.
Hence, in any case there will be an increase in  the directed
electron energy.  The escaping electrons imply that a smaller amount
of them are available to shield ion perturbations and the mode
should be even more growing. The electron parallel velocity  in a
time-varying parallel wave-electric field $E_0 \cos (k_z z -
\omega_r t)$ is
 \[
v(t)= v_0-\frac{e E_z}{m_e(k_z v_0 - \omega_r)} \left\{ \sin[k_zz_0
+ (k_z v_0 - \omega_r) t] \right.
\]
\be
\left.
 - \sin(k_zz_0)\right\}. \label{dv} \ee
Here $v_0$ and $z_0$ are the starting electron velocity and position
in the parallel direction, respectively. A strong
acceleration will take place, in particular for
particles close to resonance $v_0=\omega_r/k_z=0.1 \vte$. The
non-relativistic energy radiated by an electron  decelerated in the
time-varying wave electric field is given by $\Delta \Sigma/\Delta
t=  e^2 (\Delta v/\Delta t)^2/(6\pi \varepsilon_0  c^3)$. Here,
$\Delta t\simeq \tau_{osc}/4$, where $\tau_{osc}=2 \pi/(k_z v_0 -
\omega_r)$. It is seen that for particles close to resonance, the
oscillation period $\tau_{osc}$ becomes very large and  the same
holds for the particle velocity. So the previously  described
process of stochastic   heating will be accompanied by a large
directed acceleration and by radiation as well.

For plasma $\beta$ above the electron/ion mass ratio the drift wave
 couples with the Alfv\'{e}n wave. This coupling is
given by \cite{v3,weil}  $\omega^3 - \omega^2
(\omega_{*e}+\omega_{*i}) + \omega [\omega_{*e}\omega_{*i} - k_z^2
c_a^2/(1+ k_y^2 \rho_i^2) - k_y^2 k_z^2 c_a^2 (\rho_i^2 + \rho_s^2)]
+ k_z^2 c_a^2 \omega_{*e}/(1+ k_y^2 \rho_i^2) =0$,
$c_a^2=B_0^2/(\mu_0 n_0 m_i)$. It describes the drift wave and two
 Alfv\'{e}n waves.
       The electromagnetic effects in  RERs are frequently
       (but not always) observed. However, even in the presence of coupling,  the previous analysis
       related to the energy release will not change considerably.
       This can be checked by solving the above dispersion equation for
       the coupled modes and comparing to previous results. As
       example, for   $\lambda_y=3$ m, $s=10^3$  and other parameters
       unchanged,  the drift wave frequency from Eq. (\ref{k1}) is $0.21$ Hz, while from
       the coupled mode it is $0.25$ Hz.

\vspace{0.5cm}

\noindent{\bf III. \,\,\, SUMMARY}

\vspace{0.5cm}

The model and results presented here, based on the drift wave theory,
 yield an alternative description of some rapid energy releases in the solar corona. Because of a high temperature and a low number density,
collisions  are rare in coronal magnetic structures, hence the kinetic description of the drift wave is the most
appropriate. In such a description the drift wave is strongly growing and its growth-rate appears as a purely kinetic
effect.

Using parameters typical  for inner corona we have shown that such a growing  drift wave can lead to  stochastic
heating of ions,$^{20}$ with the heating rate so large that can be used for the description of nano-flares and smaller
energy releases. We have also shown  that there should be an  acceleration of plasma particles, primarily electrons, by
growing drift waves in solar corona.
 Such an acceleration and heating of coronal plasma has been discussed in many studies in the literature in the past,$^{31-33}$ yet not within  the  drift wave theory.
Our  analysis comprises the electrostatic limit, which may be appropriate at least in some cases because observations
in the past have shown that not all rapid energy releases are associated with a measurable change in the magnetic field
topology. However, using the well known theory the magnetic effects can be included and they would  lead to the
coupling between the drift and Alfv\'{e}n waves, and this limit is also briefly discussed.

The model implies the density gradients in the background plasma, that are expected within coronal magnetic structures.
Since the growth of the mode is on the account of the energy stored in this gradient, it is expected that the
background is simultaneously changed in the process of the instability. Simulation in the past have shown this.$^{34}$
As a result the  energy release should be below the value obtained in the present study. A  detailed description of
this feedback effect can be performed only numerically, yet this is beyond the scope of the present work.

\vspace{1cm}

\paragraph{Acknowledgements:}

These results were obtained in the framework of the projects GOA/2009-009 (K.U.Leuven), G.0304.07 (FWO-Vlaanderen) and
C~90347 (ESA Prodex 9). Financial support by the European Commission through the SOLAIRE Network (MTRN-CT-2006-035484)
is gratefully acknowledged.

\pagebreak

\end{document}